%Paper: gr-qc/9209004
%From: STEP8030@bureau.ucc.ie
%Date: Fri, 11 Sep 1992 10:41 GMT

\magnification 1200
\baselineskip=16pt
\centerline{\bf A Quasilocal Test of the Finiteness of the
Universe} \vskip 2cm
\centerline{Edward Malec$^{+}$ and Niall \'O Murchadha$^*$           }
\vskip 2cm
\centerline{$^{+}$ Instytut Fizyki, Uniwersytet Jagiello\'nski
30-064 Krak\'ow, Reymonta 4, Poland }
\centerline{$^*$ Physics Department, University College,
Cork, Ireland}
\vskip 2cm
{\bf The Cosmological Principle states that the universe is both homogeneous
and isotropic. This, alone, is not enough to specify the global geometry of
the spacetime. If we were able to measure both the Hubble constant and the
energy density we could determine whether the universe is open or closed.
Unfortunately, while some agreement exists on the value of the Hubble constant,
the question of the energy density seems quite intractable. This Letter
describes a possible way of avoiding this difficulty and shows that if one
could measure the rate at which light-rays emerging from a surface expand, one
might well be able to deduce whether the universe is closed.}
\vskip 0.8cm
There is a general belief that the universe is, on average, both isotropic and
homogeneous. This means that at any instant of time we need only three
parameters to describe the large-scale properties of the system. These are the
`radius', $a$, of the universe, the Hubble constant, $H$, and the average
density, $\rho_0$, of the matter-content of the universe. These are not
independent, but are related by $${k \over a^2} + H^2 = {8 \pi \rho_0 \over 3}.
\eqno(1)$$ $H, a$ and $\rho_0$ are time-dependent constants; the equation of
state of the matter determines their rate-of-change. The constant $k$ in
eqn.(1)
does not change with time, rather it stays fixed at either +1, 0 or -1,
depending on whether the universe is closed, flat or open (hyperbolic).
Deciding
which of these values is correct is probably the single most important question
in cosmology today. Obviously, if one knew $H$ and $\rho_0$, one could
immediately calculate $k$ and $a$.

There is agreement that one can determine (at least approximately) the value
of the Hubble constant, $H$. The major difficulty is in finding the value of
$\rho_0$ and it has been suggested that this quantity may be dominated by
material which is hard to detect, i.e., WIMPs, dust, massive neutrinos,
tiny black holes, \dots. It may well be that the energy density in our local
neighbourhood differs significantly from the average taken over the whole
universe. In this Letter we wish to suggest a new approach to determining the
global structure of the universe which does not depend on evaluating
$\rho_0$.

Consider any two-dimensional spatial surface $S$ in a spacetime.  Now consider
light-rays which emerge perpendicular to this surface. The outgoing
light-rays will (in general) diverge and one can compute the expansion
$\theta$ of these light-rays by measuring the fractional change of the area of
a
little element of $S$ as it is dragged along these light rays. This quantity
plays a key role in one of the singularity theorems of general relativity; if
$\theta < 0$ on $S$, then $S$ is called an (outer) trapped surface and it
indicates the presence of a future singularity [1, 2]. We have been involved in
an ongoing investigation to discover when such trapped surfaces would form in a
cosmological context. As part of this analysis, we have discovered a very
simple relationship which may allow us to use the value of this expansion to
determine whether the universe is closed.

The expansion in a null directon can be regarded as a combination of the
expansion in a time direction and one in a spatial direction. If the spatial
3-slice is closed the area of any `expanding' 2-surface will eventually start
diminishing. We therefore expect that the expansion $\theta$ should contain
information not only about the local geometry but also about the global
topology.

We wish to consider a spherically symmetric universe (which we assume to be
isotropic but not necessarily homogeneous) and we choose a time-slice which is
the ``Hubble-time'' slice through this universe, i.e., that the Hubble
`constant' is in fact constant. We assume that this slice respects the
spherical symmetry (this is probably automatic). We further assume that the
matter-field is instantaneously at rest. Finally, we assume that the matter
density fluctuates (not necessarily by small amounts) about some average value,
$\rho_0$, so that on some large scale the slice looks like a standard slice
through one of the standard Friedmann universes. Now consider any spherically
symmetric surface S.   Consider the expansion $\theta |_S$ of the outgoing
null-rays from S. We have derived the following inequality [3, 4, 5]
 $$\theta |_S > {4\pi L \over A} + 2H - {3kV
\over A a^2} + {\Delta M \over A}, \eqno(2)$$
 where  L is the proper radius of S, A is the area of S, V is the volume inside
S and $$\Delta M = \int_V (\rho - \rho_0) dv \eqno(3)$$ is the integral of the
excess energy-density inside S, where $\rho$ is the true (nonconstant) energy
density. Inequality (2) has been shown to hold for each of the three values of
$k$, the three choices of the global topology.

Let us now choose S large enough that the fluctuations in $\rho$
average out. In this case $\Delta M $ vanishes and inequality (2) simplifies to
 $$\theta |_S > {4\pi L \over A} + 2H - {3kV \over A a^2}. \eqno(4)$$
Say we find a large spherical surface S which satisfies
$$\theta |_S < {4 \pi L \over A} + 2H . \eqno(5)$$
Inequality (5) is only compatible with (4) when $k$ = +1. Hence we can deduce
that the universe is closed.

The important thing to notice is that both $\rho_0$ and $a$ play no role in
(5), so, as promised, we have found an inequality which may be used to
determine whether the universe is open or closed, without ever measuring
either the local or global energy density. Further, inequality (5) is not
vacuous, surfaces satisfying this condition can be found in any closed
spherical cosmology.

It is clear that this inequality can only be worked one way,
we cannot use it to find a condition that guarantees that the universe is
flat or open. This is in keeping with the observation, first made by Einstein
[6], that, while it is possible to demonstrate conclusively that the universe
is
closed, it is essentially impossible to prove the converse.

 We stress that that the key inequality (2) is not in any sense a perturbative
result, it holds true even for large deviations of $\rho$ from the average.
 Further, while we used spherical symmetry to derive our results,
it is clear that the inequalities are stable under non-spherical perturbations.
The key assumption we make is that one can determine a cosmological scale, a
measure of the spatial extent of the fluctuations of the matter, because we
need to find a large enough surface so that the fluctuations average out. We
claim that this is not the same as determining the actual matter distribution;
all we need to assume is that the dark matter drags the visible matter with it.
\vskip 0.5cm
{\bf Acknowledgements.} We would like to thank the Physics Department,
University College Cork for inviting one of us (EM) to Cork, where  part of
this work was done.  This work has been  supported by the  Polish Government
Grant no. PB 2526/2/92.
\vskip 1cm
\centerline{\bf References.}
\vskip 0.5cm
[1] R. Penrose, {\it Phys. Rev. Lett.} {\bf 14}, 57 (1965);{\it Techniques
of Differential Topology in Relativity} (Soc. Ind. Appl. Math. 1972).\par
[2] S. W. Hawking, G.F.R. Ellis, {\it The large scale structure of
space-time} (Cambridge University Press, Cambridge 1973). \par
[3] U. Brauer, E. Malec, {\it Phys. Rev}. {\bf D45}, R1836 (1992).\par
[4] E. Malec, N.\'O Murchadha, submitted to {\it Phys. Rev.}\par
[5] U. Brauer, E. Malec, N. \'O Murchadha, to be published. \par
[6] A. Einstein {\it The Meaning of Relativity,} fifth edition (Princeton
University Press, Princeton, 1955).

\end